\documentclass[doublecol]{epl2}

\title{Winding numbers of phase transition points for one-dimensional topological systems}
\author{Linhu Li\inst{1,2} Chao Yang\inst{1} Shu Chen\inst{1,3}}

\shortauthor{L. Li, C. Yang, S. Chen}

\institute{
  \inst{1} Beijing National Laboratory for Condensed Matter
Physics, Institute of Physics, Chinese Academy of Sciences, Beijing
100190, China\\
  \inst{2} Beijing Computational Science Research Center, Beijing, 100089, China\\
  \inst{3} Collaborative Innovation Center of Quantum Matter, Beijing, China
}


\abstract{We study topological properties of phase transition points of one-dimensional topological quantum phase transitions by assigning winding numbers defined on closed circles around the gap closing points in the parameter space of momentum and a transition driving parameter, which overcomes the problem of ill definition of winding numbers on the transition points. By applying our scheme to the extended Kitaev model and extended Su-Schrieffer-Heeger model, we demonstrate that the topological phase transition can be well characterized by winding numbers of transition points, which reflect the change of the winding number of topologically different phases across the phase transition points.}

\begin{document}

\maketitle

\section{Introduction}
Due to the rapid progress in the study of topological insulators and superconductors\cite{review1,review2}, theoretical studies of topological phases and phase transitions have attracted great interest\cite{Volvik,Shen-book,Murakami,Ma,NiuY,SongJ,Viyuela,Castro,YangC,SongZ}. Unlike conventional quantum phase transitions (QPTs)\cite{QPT}, topological QPTs do not accompany symmetry breaking, but involve the change of ground-state topological properties. Different topological states can be classified by quantized topological invariants, while the different phases in conventional QPTs are distinguished by order parameters, which generally take continuous values. Conventional QPTs can be classified by singularity properties of ground-state energy at the phase transition point, but this method does not unveil the topological properties of the phase transition point, and can not distinguish a topological phase transition from a conventional one. On the other hand, topological invariants are usually defined for quantum states protected by nonzero energy gaps, hence they are ill defined at topological phase transition points, where the band gaps close at some specific points in the Brillouin zone (BZ). Nevertheless, it was recently indicated that the topological properties of phase transition points can be characterized by introducing topological invariants defined on closed curves surrounding the transition points in the enlarged parameter space \cite{TQPT}, which provides a scheme to study the topological properties of phase transition points.

Since its discovery \cite{Berry}, the Berry phase has played an important role in many quantum phenomena \cite{Niu}. A well-known example is that the quantized transversal conductivity in the quantum Hall effect can be measured by the Chern number, i.e., the integral of the Berry curvature over the two-dimensional BZ. The topological properties of one-dimensional (1D) systems can be classified by the Berry phase across the BZ (i.e. the Zak phase\cite{zak}), which can be experimentally measured in 1D optical lattices \cite{Berry_measurement}. The Zak phase $\gamma$ takes the value of $\pi$ with a modulus of $2\pi$ for topologically nontrivial systems which hold a pair of degenerate edge states\cite{Ryu}, while there are no edge states in topologically trivial systems with $\gamma=0$. However, for Z-type 1D topological systems, there may exist different topological phases with more than one pair of degenerate edge states \cite{SongJ,NiuY}, and can not be fully classified by the Zak phase. For 1D topological systems, the winding number is a more convenient quantity to characterize the topological properties of the Z-type systems, but it also suffers from the problem of ill definition on topological phase transition points.

In this letter, we first give a definition of the winding number for the phase transition point in the enlarged parameter space by treating the phase driving parameter as an additional parameter besides the momentum. Such a definition overcomes the ill-definition problem at the transition point as it is defined on a closed detour path surrounding the gapless transition point and enables us to study the topological property of the transition point. Via some straightforward arguments, we find a simple relation between the defined winding number for the transition point and the winding numbers on two sides of the transition point defined in the BZ, which explains why it can be used to classify various topological phase transitions. To exemplify our theory, we then apply our scheme to study the extended Kitaev's p-wave superconductor model and the extended Su-Schrieffer-Heeger (SSH) model which both exhibit rich phase diagrams.

\section{Winding number for phase transition point}
We consider a general two-band 1D $Z$-type topological system, whose Hamiltonian in momentum space only contains two of the three Pauli's matrices, and can be written in the form of
\begin{equation}
H(k,\eta)=h_0 I + h_x \sigma_x + h_y \sigma_y
\end{equation}
after some rotations, with $I$ the unit matrix and $\sigma$ the Pauli's matrices. Here $\eta$ represents a phase transition driving parameter, $k$ is the 1D momentum, and $h_0$, $h_x$, $h_y$ are generally functions of parameters $\eta$ and $k$.
The winding number for 1D systems can be associated to the Zak phase
\begin{eqnarray}
\gamma=\int_{0}^{2\pi}dk\langle u_k|i\partial_k|u_k\rangle,
\label{zak}
\end{eqnarray}
where $u_k$ denote the occupied Bloch states of the Hamiltonian. The eigenstate for the lower band then has the form of $u_{k}=\frac{1}{\sqrt{2}}(\frac{h_x-ih_y}{\sqrt{h_x^2+h_y^2}},-1)^T$, thus the Zak phase
\begin{eqnarray}
\gamma&=&\frac{1}{2}\int_{-\pi}^{\pi}dk \frac{h_x \partial_k h_y-h_y \partial_k h_x}{h_x^2+h_y^2}\\
&=&\nu \pi,
\end{eqnarray}
where
\begin{eqnarray}
\nu=\frac{1}{2\pi}\oint_c \frac{h_x dh_y-h_y dh_x}{h_x^2+h_y^2} \label{nu}
\end{eqnarray}
is the winding number of the Hamiltonian, and $c$ is a close loop with $k$ varying from $0$ to $2\pi$. The winding number describes the total number of times that the Hamiltonian travels counterclockwise around the origin. Despite that the Zak phase only takes $0$ or $\pi$ with a modulus of $2\pi$, the winding number can be any integer, and indicates different topological phases. From Eq.(\ref{nu}), one can see that the definition of $\nu$ is invalid whenever $|h| = \sqrt{h_x^2+h_y^2} = 0 $. As the energy spectrum of $H(k)$ is given by $E(k) = h_0 \pm \sqrt{h_x^2+h_y^2}$, it is obvious that the winding number is ill defined at the phase transition point $\eta_0$, where the gap between two bands closes and one has $h_x(k_0, \eta_0)=h_y(k_0, \eta_0)=0$.
\begin{figure}
\includegraphics[width=0.8\linewidth]{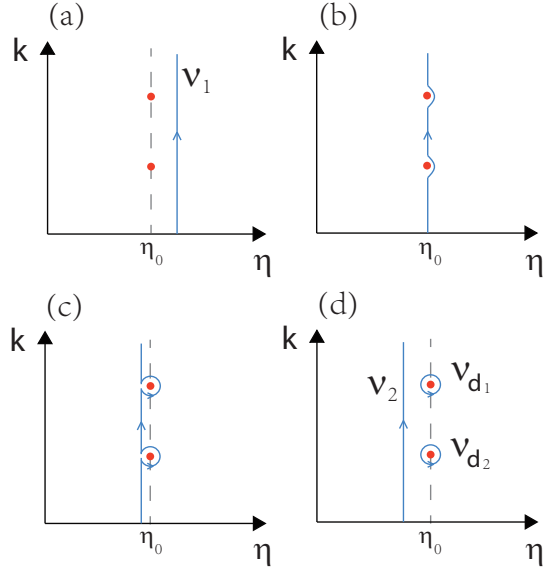}
\caption{(Color online) The sketch of the integral path of the winding number. $\eta_0$ is the phase transition point, and the red dots stand for the degenerate points at which $|h(k)|=0$. $\nu_1$ ($\nu_2$) is the winding number of the system on the right (left) side of the dashed line with $\eta=\eta_0$. The integral path is smoothly moved from one topological phase to another without passing the degenerate points.} \label{fig1}
\end{figure}

To characterize the topological properties of the phase transition points, we define the winding number $\nu_d$ on a circle around the gap closing point in the parameter space of $k$ and $\eta$. Following reference \cite{TQPT}, we set $k=A\sin{\theta}+k_0$ and $\eta=A\cos{\theta}+\eta_0$, with $\theta$ the varying angle, $A$ the radius of the circle, and $(k_0,~\eta_0)$ the gap closing point in the parameter space of $k$ and $\eta$. Hence the Hamiltonian around the circle can be represented as $h(\theta)$, and the winding number is defined as
\begin{eqnarray}
\nu_d=\frac{1}{2\pi}\oint_{c'} \frac{h_x dh_y-h_y dh_x}{h_x^2+h_y^2}\label{winding_d}
\end{eqnarray}
with $c'$ a close loop from $\theta=0$ to $\theta=2\pi$.

Setting $h_x/|h|=\cos{\alpha}$, one can easily have
\begin{eqnarray}
\nu~=\frac{1}{2\pi}\oint_{c} d\alpha, \\
\nu_d=\frac{1}{2\pi}\oint_{c'} d\alpha,
\end{eqnarray}
which directly indicates the times that the Hamiltonian goes around the origin along the close loop $c$ (or $c'$).
Given a quantum state in the regime of $\eta > \eta_0$, the corresponding winding number $\nu_1$ is only defined on the momentum $k$ space. As there are no singular points in the space of ($k$, $\eta$) except of the gapless transition points schematically marked by dots in Fig.\ref{fig1}, one can analytically continue the integral path $c$ to the two-dimensional parameter space.No matter how we change the integral path, the winding number is unchanged as long as the Hamiltonian does not cross the degenerate points, at which $|h(k)|=0$. As schematically displayed in Fig.\ref{fig1}, one can smoothly move the integral path pass the phase transition point. Such a continuous change of integral path does not change the value of the integral, but leaves close loops around each degenerate point and a close path only on the momentum space (as shown in Fig.\ref{fig1} (d)). Effectively, this process leads to
\begin{eqnarray}
\sum_i \nu_{d_i}= \nu_1 - \nu_2,
\end{eqnarray}
where $\nu_1$ and $\nu_2$ denote winding numbers of different phases on two sides of the transition point, and $\nu_{d_i}$ denotes the winding number of the $i$-th degenerate points.
The above equation indicates clearly that the change of winding number $\nu$ across the phase transition point equals to the summation of winding number $\nu_d$ of each degenerate points.

\section{Extended Kitaev model}
To give a concrete example and verify our scheme, first we consider the extended version of the Kitaev model \cite{Kitaev} by adding next-nearest neighboring hopping and pairing terms, descried by the Hamiltonian
\begin{eqnarray}
H&=&\sum_i t_1\hat{c}^{\dagger}_{i}\hat{c}_{i+1}+t_2\hat{c}^{\dagger}_{i}\hat{c}_{i+2}+h.c.\nonumber\\
&&+\Delta_1\hat{c}_{i}\hat{c}_{i+1}+\Delta_2\hat{c}_{i}\hat{c}_{i+2}+h.c.-2\mu\hat{c}^{\dagger}_{i}\hat{c}_{i},
\end{eqnarray}
where $\hat{c}^\dagger_i$ ($\hat{c}_i$) is the creation (annihilation) operator of fermions at the {\it i}-th site.

After the Fourier transformation, it takes the form of
\begin{equation}
H = \psi^{\dagger}_k H(k) \psi_k ,
\end{equation}
where $\psi^{\dagger}_k = (c_{k},c^{\dagger}_{-k})$ and
\begin{eqnarray}
H(k)=\left(
\begin{array}{cc}
h_z &-ih_y\\
ih_y &-h_z
\end{array}\right),\label{H_Kitaev}
\end{eqnarray}
with $h_y=\Delta_1\sin{k}+\Delta_2\sin{2k}$ and $h_z=-t_1\cos{k}-t_2\cos{2k}+\mu$. The eigenvalues are given by
\begin{eqnarray}
E(k) = \pm |h(k)|= \pm \sqrt{h_y^2 + h_z^2 }.
\end{eqnarray}
This model is related to the extended quantum Ising model with additional three-body interaction by Jordan-Wigner transformation \cite{NiuY,SongZ}. For simplicity, in the following discussion we shall set $\Delta_1=t_1$ and $\Delta_2=t_2$ with $t_{1,2}$ taking real numbers \cite{NiuY}. In general, according to the the ten-fold-way classification\cite{classification}, the model with complex $\Delta_{1,2}$ belongs to the D class, which shall be characterized by a $Z_2$ quantity. However, for the case with real  $\Delta_{1,2}$ considered in the present work, due to the absence of $\sigma_x$ in the Hamiltonian, this model belongs to the Z-type topological system which can be characterized by the winding number.
\begin{figure}
\includegraphics[width=0.9\linewidth]{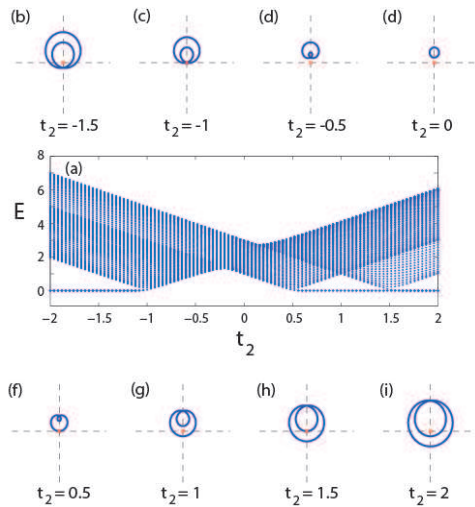}
\caption{(Color online) The energy spectrum under OBC versus $t_2$ and the winding of the Hamiltonian with different $t_2$ of the extended Kitaev model. The parameters of (a) are $\mu=1$, $t_1=\Delta_1=0.5$ and $\Delta_2=t_2$. The times that the Hamiltonian goes around the origin indicate the winding number of the corresponding topological phase. In (c), (f) and (h), the Hamiltonian goes through the origin, and the winding number is ill defined.} \label{fig2}
\end{figure}
\begin{figure}
\includegraphics[width=0.8\linewidth]{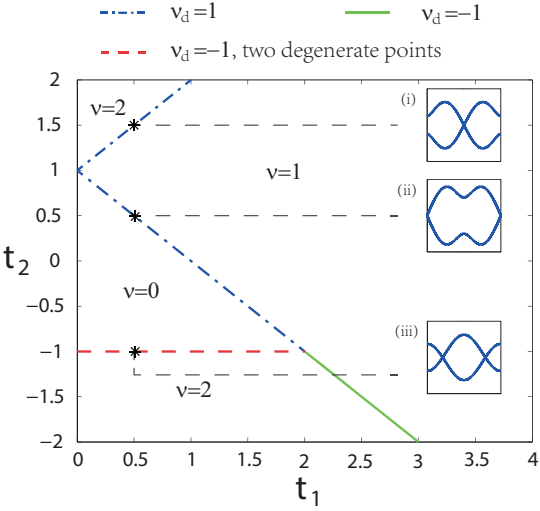}
\caption{(Color online) The phase diagram of the extended Kitaev model with the winding number $\nu_d$ on the phase boundary being marked. Here we have taken $\mu=1$, $t_1=\Delta_1$, and $t_2=\Delta_2$. There are two degenerate points in the BZ for the system on the red dashed line, and the summation of $\nu_d$ is $-2$. Inserts show the energy spectrum in momentum space for systems on the phase transition points labeled by stars. } \label{fig3}
\end{figure}

Since the energy gap must close at phase transition points, we can determine the phase boundaries of the extended Kitaev model by the gap closing condition $|h(k)|=0$, which yields $(k=0,~t_1+t_2-\mu=0)$, $(k=\pi,~-t_1+t_2-\mu=0)$ or $(k=\pm\arccos{(-t_1/2t_2)},~t_2=-\mu)$. Under open boundary condition (OBC), this model may hold different number (0, 1 or 2) of Majorana zero modes at each end. In Fig.\ref{fig2}(a), we show the energy spectrum under OBC with $\mu=1$, $t_1=\Delta_1=0.5$ and $\Delta_2=t_2$. Here we only display the upper branch of the spectrum with $E \geq 0$ due to the chiral symmetry of the spectrum.  With increasing of $t_2$, we can see that the system goes through different phases labeled by the appearance and the disappearance of zero modes. Notice that there are two pairs of Majorana zero modes in the regime of $t_2<-1$ or $t_2>1.5$, and only one pair in the regime of $0.5<t_2<1.5$.

Next we calculate the topological invariant of this model. we rotate the Hamiltonian to the x-y plane with $h_y\rightarrow h_x,~h_z\rightarrow h_y$ to coincide with the definition of the winding number in previous section. In Fig.\ref{fig2}(b)-(i), we illustrate the winding of the Hamiltonian with different $t_2$. When the momentum $k$ varies from $0$ to $2\pi$, the curve of the Hamiltonian may enclose the origin twice, once, or not enclose the origin, which correspond to $\nu=2$, $1$, $0$, respectively. The number of times that the Hamiltonian goes around the origin indicates the number of Majorana zero modes at each end of an open chain. However, at the phase transition points (as shown in Fig.\ref{fig2}(c), (f) and (h)), the curve crosses the origin, and the winding number is ill defined due to $|h|=0$ at the origin.

To characterize the topological properties of the phase transition points, we calculate the winding number $\nu_d$ defined in the previous section by choosing $t_2$ as the phase transition driving parameter, and demonstrate the phase diagram of this model with $\mu=1$ in Fig.\ref{fig3}. We can see that the summation of $\nu_d$ equals to the change of $\nu$ across the phase transition point. The inserts (i), (ii) and (iii) in Fig.\ref{fig3} show the energy spectrum of phase transition points corresponding to Fig \ref{fig2} (c), (f) and (h), respectively. There is only one degenerate point with $\nu_d=1$ in (ii) and (iii), but two degenerate points emerge in (iii) with $\nu_d=-1$ for each of them. While the change of $\nu$ on two sides of the dashed-dotted lines is $1$, the change of $\nu$ across the dashed line is $-2$, which corresponds to the summation of $\nu_d$ of these two degenerate points.

\section{Extended SSH model}
Next we consider an extended SSH model \cite{SongJ}, which takes the form of
\begin{equation}
H = \psi^{\dagger}_k H(k) \psi_k ,
\end{equation}
where $\psi^{\dagger}_k = (c^{\dagger}_{k,A},c^{\dagger}_{k,B})$ and
\begin{eqnarray}
H(k) =
\left(
\begin{array}{cc}
0 & t_1 + t_2 e^{-ik}+ t' e^{-2ik} \\
t_1 + t_2 e^{ik}+ t' e^{2ik} &
0%
\end{array}%
\right) .
\end{eqnarray}
Alternatively, $H(k)$ can be expressed in the form $H(k)=h_x\sigma_x+h_y\sigma_y$, with $h_x =  t_1 + t_2 \cos k +  t' \cos 2k  $ and $h_y=  t_2 \sin k  + t' \sin 2k $. Diagonalizing the Hamiltonian, we get the eigenvalues
\begin{eqnarray}
E(k) = \pm |h(k)|= \pm \sqrt{h_x^2 + h_y^2 } .
\end{eqnarray}
When $t'=0$, the model reduces to the SSH model \cite{SSH} of the BDI class,
which can be described by the Hamiltonian:
\begin{eqnarray}
H=\sum_i [t_1 \hat{c}^{\dagger}_{A,i}\hat{c}_{B,i}+t_2 \hat{c}^{\dagger}_{A,i+1}\hat{c}_{B,i}]+h.c.,
\end{eqnarray}
where $c^{\dagger}_{A(B),i}$ is the creation operator of fermion on
$i$-th A (or B) sublattice.  It is
well known that this model has two topologically distinct phases. For convenience, we set $t_1=t(1-\delta)$, $t_2=t(1+\delta)$ and take $t=1$ as the energy unit. When $\delta>0$, the system belongs to a topological nontrivial phase with the winding number $\nu=1$, and supports two degenerate zero-mode edge states for the model with OBC. When $\delta<0$, the system has $\nu=0$ and the edge states disappear.
\begin{figure}
\includegraphics[width=0.8\linewidth]{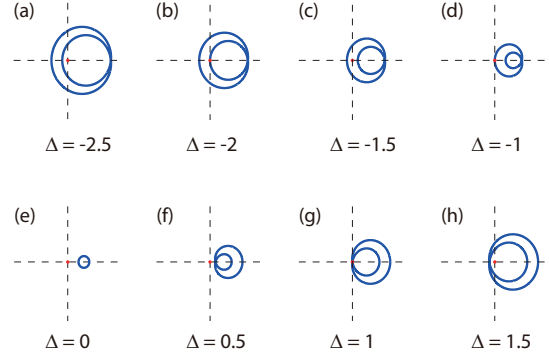}
\caption{(Color online) The winding of the Hamiltonian of the extended SSH model with different $\Delta$ and $\delta=-0.5$. The times that the Hamiltonian goes around the origin indicate the winding number of the corresponding topological phase. In (b), (d) and (g), the Hamiltonian goes through the origin, and the winding number is ill defined.} \label{fig4}
\end{figure}
\begin{figure}
\includegraphics[width=0.8\linewidth]{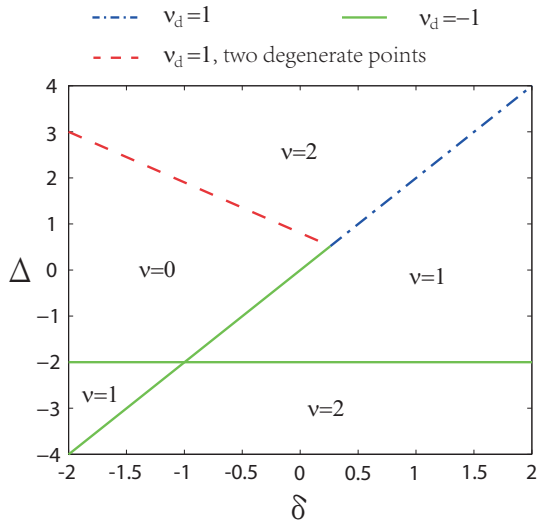}
\caption{(Color online) The phase diagram of the extended SSH model with $\nu_d$ on the phase boundary being marked. There are two degenerate points in the BZ for the system on the red dashed line, and the summation of $\nu_d$ is $2$.} \label{fig5}
\end{figure}

When $t'\neq0$, the system still belongs to the BDI class as the time reversal symmetry and the chiral symmetry are preserved, but the topological phases of the extended SSH model are enriched. Defining $t'=\Delta t$, the gap closing conditions are given by $(k=0,~\Delta=-2)$, $(k=\pi, \Delta=2\delta)$ or $(k=\pm\arccos{[-(1+\delta)/2\Delta]},~\delta+\Delta=1)$. By fixing $\delta=-0.5$, we show the winding pattern of the Hamiltonian with different parameters in Fig.\ref{fig4}. It is clear that the winding pattern shows different geometrical structure in various parameter regimes. For the Fig.\ref{fig4}(a) and (h), we can see that the curve goes around the origin twice, which corresponds to the winding number $\nu=2$. Similarly, we have $\nu=1$ corresponding to Fig.\ref{fig4}(c) and $\nu=0$ corresponding to Fig.\ref{fig4}(e) and (f). On the other hand, Fig.\ref{fig4} (b), (d) and (g) correspond to some phase transition points, at which the Hamiltonian goes through the origin and the winding number $\nu$ is ill defined.

Next we display the phase diagram of the extended SSH model in Fig.\ref{fig5} with topologically different phases characterized by different values of the winding number $\nu$. By choosing $\Delta$ as the driving parameter, we can calculate the winding number $\nu_d$ for the phase transition points. As shown in Fig.\ref{fig5}, we can see that $\nu_d$ equals to the change of the winding number $\nu$ across the phase transition point. The state at the red dashed line with $\delta+\Delta=1$ is a 1D semimetal with band degenerate points at $k=\pm\arccos{[-(1+\delta)/2\Delta]}$, while the ones at other phase transition boundaries have only one degenerate point in the BZ.

\section{Summary}
In summary, we have studied the topological properties of phase transition points of 1D topological superconductor and insulator systems. We characterized different topological phases with the winding number, and defined the winding number for phase transition points in the parameter space of system's momentum and a transition driving parameter. We studied several 1D topological models with high winding number, and demonstrated that the topological phase transitions can be characterized by the introduced winding number around the transition point, the summation of which equals to the change of the winding numbers of different phases across the transition point. Our theory provides a way to classify different topological phase transitions by directly studying the properties of the phase transition point and can be applied to study other 1D Z-type topological systems.

\acknowledgments
The work is supported by NSFC under Grants No. 11425419, No. 11374354 and No. 11174360.


\begin{thebibliography}{0}

\bibitem{review1}
\Name{Hasan M. Z. \and Kane C. L.}\REVIEW{Rev. Mod. Phys.}{82}{2010}{3045}.

\bibitem{review2}
\Name{Qi X.-L. \and Zhang S.-C.}\REVIEW{Rev. Mod. Phys.}{83}{2011}{1057}.

\bibitem{Volvik} \Name{Volovik  G. E.} \emph{The Universe in a Helium Droplet} (Oxford University Press,  Oxford, 2003).

\bibitem{Shen-book} \Name{Shen S.-Q.} \emph{Topological Insulators} (Springer-Verlag, Hei-
delberg, 2013).

\bibitem{Murakami} \Name{Murakami S} \REVIEW{New J. Phys.} {9} {2007} {356}.

\bibitem{Ma} \Name{Ma Y. Q. , Chen S., Fan H., and Liu W. M. } \REVIEW{Phys. Rev. B} {81} {2010}
{245129}.

\bibitem{NiuY} \Name{Niu Y., Chung S. B., Hsu C.-H., Mandal I., Raghu S. and Chakravarty S.} \REVIEW{Phys. Rew. B}{85}{2012}{035110}.

\bibitem{SongJ} \Name{Song J. and Prodan E.} \REVIEW{Phys. Rev. B}{89}{2014}{224203}.

\bibitem{Viyuela}
\Name{Viyuela O., Rivas A. and Martin-Delgado M. A.} \REVIEW{Phys. Rev. Lett.} {112} {2014} {130401}.

\bibitem{Castro}
\Name{Sacramento P. D., Ara¨²jo M. A. N. and Castro E. V.} \REVIEW{Europhys. Lett}{105}{2014}{37011}.

\bibitem{YangC}
\Name{Yang C., Guo H., Fu L.-B. and Chen S.} \REVIEW{Phys. Rev. B}{91}{2015} {125132}.

\bibitem{SongZ} \Name{Zhang G. and Song Z.} arXiv:1504.00256.

\bibitem{QPT}
  \Name{Sachdev S.}
  \Book{Quantum Phase Transitions}
  \Publ{Cambridge University Press, Cambridge, England,}
  \Year{1999}

\bibitem{TQPT} \Name{Li L., Chen S.} ArXiv:1503.04959.

\bibitem{Berry} \Name{Berry M. V.} \REVIEW{Proc. R. Soc. London, Ser. A}{392}{1984}{45}.

\bibitem{Niu} \Name{ Xiao D., Chang M. C. and Niu Q.} \REVIEW{Rev. Mod. Phys.} {82} {2010}
{1959}.

\bibitem{zak} \Name{Zak J.} \REVIEW{Phys. Rev. Lett.}{62}{1989}{2747}.

\bibitem{Berry_measurement} \Name{Atala M., Aidelsburger M., Barreiro J. T., Abanin D., Kitagawa T., Demler E., Bloch I.} \REVIEW{Nat. Phys.}{9}{2013}{795}.

\bibitem{Ryu} \Name{Ryu S. and Hatsugai Y.} \REVIEW{Phys. Rev. Lett.}{89}{2002}{077002}.

\bibitem{Kitaev} \Name{Kitaev A. Y.} \REVIEW{Phys. Usp.}{44}{2001}{131}.

\bibitem{classification}
\Name{Altland A. and Zirnbauer M.} \REVIEW{Phys. Rev. B}{55}{1997}{1142};
\Name{Schnyder A. P., S. Ryu, Furusaki A. and Ludwig A. W. W.} \REVIEW{Phys. Rev. B}{78}{2008}{195125};
\Name{Kitaev A.} \REVIEW{AIP Conf. Proc}{1134}{2009}{22}.

\bibitem{SSH} \Name{Su W. P., Schrieffer J. R. and Heeger A. J.} \REVIEW{Phys. Rev. Lett.}{1698}{1979}{42}.

\end{thebibliography}
\end{document}